%% file: main.tex
\documentclass[conference]{IEEEtran}
\usepackage[top=0.75in, bottom=1.1in, left=0.625in, right=0.625in]{geometry}
\IEEEoverridecommandlockouts
\usepackage{cite}
\usepackage{amsmath}   % usually already loaded
\usepackage{bbm}
\newcommand{\ind}{\mathbbm{1}}
\usepackage{amsmath,amssymb,amsfonts}
\usepackage{algorithmic}
\usepackage{graphicx}
\usepackage{textcomp}
\usepackage{subcaption}
\usepackage{caption}
\usepackage{xcolor}
\def\BibTeX{{\rm B\kern-.05em{\sc i\kern-.025em b}\kern-.08em
    T\kern-.1667em\lower.7ex\hbox{E}\kern-.125emX}}
\usepackage[acronym]{glossaries}
\usepackage{booktabs}
\usepackage{array}
\usepackage{multirow}
\usepackage{multicol}
\usepackage[normalem]{ulem}
\usepackage{url}
\input{acronyms}

\begin{document}

\title{ELC: \underline{E}vidential \underline{L}ifelong \underline{C}lassifier for Uncertainty Aware Radar Pulse Classification\\
\thanks{This study was supported by the EME Hub, funded by Defence Science and Technology Laboratory (Dstl).}
}

\author{\IEEEauthorblockN{Mohamed Rabie}
\IEEEauthorblockA{\textit{Wolfson School of Mechanical, Electri-} \\ \textit{cal and Manufacturing Engineering} \\
\textit{Loughborough University, U.K.}\\
%Loughborough, U.K. \\
%\IEEEauthorblockA{\textit{Wolfson School of M.E.M.E.} \\
%\textit{Loughborough University, UK}\\
m.rabie@lboro.ac.uk}
\and
\IEEEauthorblockN{Chinthana Panagamuwa}
\IEEEauthorblockA{\textit{Wolfson School of Mechanical, Electri-} \\ \textit{cal and Manufacturing Engineering} \\
\textit{Loughborough University, U.K.}\\
%Loughborough, U.K. \\
%\IEEEauthorblockA{\textit{Wolfson School of M.E.M.E.} \\
%\textit{Loughborough University, UK}\\
c.j.panagamuwa@lboro.ac.uk}
\and
\IEEEauthorblockN{Konstantinos G. Kyriakopoulos}
\IEEEauthorblockA{\textit{Wolfson School of Mechanical, Electri-} \\ \textit{cal and Manufacturing Engineering} \\
\textit{Loughborough University, U.K.}\\
%Loughborough, U.K. \\
%\IEEEauthorblockA{\textit{Wolfson School of M.E.M.E.} \\
%\textit{Loughborough University, UK}\\
%Loughborough, U.K. \\
k.kyriakopoulos@lboro.ac.uk}
}

\maketitle

\begin{abstract}
Reliable radar pulse classification is essential in Electromagnetic Warfare for situational awareness and decision support. Deep Neural Networks have shown strong performance in radar pulse and RF emitter recognition; however, on their own they struggle to efficiently learn new pulses and lack mechanisms for expressing predictive confidence. 
This paper integrates Uncertainty Quantification with Lifelong Learning to address both challenges. The proposed approach is an Evidential Lifelong Classifier (ELC), which models epistemic uncertainty using evidence theory. ELC is evaluated against a Bayesian Lifelong Classifier (BLC), which quantifies uncertainty through Shannon entropy. Both integrate Learn-Prune-Share to enable continual learning of new pulses and uncertainty-based selective prediction to reject unreliable predictions. ELC and BLC are evaluated on 2 synthetic radar and 3 RF Fingerprinting datasets. Selective prediction based on evidential uncertainty improves recall by up to $46\%$ at $-\text{20~dB}$ SNR on synthetic radar pulse datasets, highlighting its effectiveness at identifying unreliable predictions in low-SNR conditions compared to BLC. These findings demonstrate that evidential uncertainty offers a strong correlation between confidence and correctness, improving the trustworthiness of ELC by allowing it to express ignorance.
% Convolutional Neural Networks have proven highly effective for \gls{rff}, i.e., identifying RF transmitters from their signals. %in terms of generalisability, robustness to noise, and feature extraction. 
% Nonetheless, conventional \gls{ml} models are limited in their ability to learn new identification tasks due to catastrophic forgetting. Another key challenge is deciding how much confidence to place in a model's prediction. 
% This paper addresses these challenges by proposing an \gls{elc}, which integrates \gls{ll} with an evidential classifier capable of expressing epistemic uncertainty. \gls{elc} enables continual learning of new waveforms without retraining, while also quantifying model confidence. \gls{elc} is evaluated on 3 \gls{rff} and 2 radar pulse datasets. Results show that \gls{elc} performs similarly, if not better, than an \gls{ll} baseline, albeit at higher computational cost. On average, \gls{elc} exhibits higher uncertainty when making incorrect predictions. This is valuable for i) determining when to trust the model, and ii) highlighting where more data or better features are needed. 
\end{abstract}

\glsresetall
\section{Introduction}
% Imperfections in \gls{rf} circuitry introduce subtle distortions to transmitted signals, which form a unique fingerprint for each transmitting device. Deep \gls{cnn} can learn to detect and classify these fingerprints. This gives rise to the concept of \gls{rff}, which aims to classify the transmitter based on its unique signal distortions. \gls{rff} has been proposed as a method for authenticating RF transmitters \cite{DLforRFF_2020}.
Reliable classification of radar waveforms is critical in \gls{ew} to aid decision making. Deep \glspl{nn} have proven effective at detecting and classifying radar emitters based on their transmissions \cite{rc_wang2017, rc_jagannath2021, rc_Huang2023}. RF signal classification, as a passive technique that relies solely on signal reception, is both energy-efficient and stealthy. 

Learning new waveforms efficiently and retaining prior knowledge are demonstrated in recent work. Transfer learning is used by exploiting prior knowledge to expedite learning of new waveforms \cite{huang2025radar_fewshot, jing2022radar_fewshot, xiao2020radar_fewshot}. Similarly, \gls{ll} is used to continually learn new classes without sacrificing performance on old classes and is being explored in the RF domain to enable continual learning of waveforms \cite{han2023_radar_il, DLforRFF_2020}.

Similar techniques are used for \gls{rff} for detection and/or classification of RF emitters such as \gls{usrp} \cite{ExposingRFF} and \gls{iot} devices \cite{DLforRFF_2020}. In both \gls{rff} and radar domains, deep \gls{cnn} have consistently outperformed traditional \gls{ml} methods that rely on expert-crafted features \cite{DLforRFF_2020, ExposingRFF}. \gls{cnn}s excel in feature extraction, generalisability across devices, robustness to noise, and classification of modulation \cite{DLforRFF_2020, rc_wang2017}. 

\gls{ll} differs from conventional \gls{st} models by enabling incremental learning of data without ``catastrophically forgetting" prior knowledge. 
% \gls{st} models, once trained, are incapable of continually learning as new data becomes available (e.g. new \gls{rf} waveforms) without ``forgetting" old data. 
Approaches for mitigating catastrophic forgetting broadly fall into two categories: Allowing new tasks to cooperatively modify the same parameters of the \gls{nn}, or partitioning the parameters into disjoint subsets for each training round \cite{cl_review2023}.
% \gls{ll} models are capable of continually learning as more data becomes available.

Another key challenge is knowing how much confidence to place in a model's predictions, especially in sensitive domains such as \gls{ew}. This is valuable for downstream decision making when incorrect predictions incur undesirable cost. \gls{uq} methods express uncertainty alongside predictions and have been shown to exhibit higher uncertainty when making incorrect predictions \cite{al_epis_uncertainty}. This is useful for identifying and rejecting unreliable predictions.
% , perhaps deferring them to an expert.
% One approach is evidential classification \cite{ecnn-ds}, which expresses epistemic uncertainty alongside predictions.

This paper proposes the integration of \gls{uq} methods within a LL context, by integrating the \gls{lps} \cite{lps} with both evidential \cite{ecnn-ds} and Bayesian classification. An \gls{elc} and a \gls{blc} are thus evaluated for continually learning new waveforms without forgetting, whilst expressing uncertainty in predictions using evidential uncertainty and Shannon entropy, respectively. These models are benchmarked against the vanilla \gls{st} and \gls{lps} approaches using a backbone ResNet \cite{resnet}. Both models use selective prediction to reject uncertain predictions. Selective prediction is evaluated on synthetic radar datasets as a function of \gls{snr}.\footnote{Code available at \url{https://github.com/mrabie9/elc}}
% however \gls{elc} shows better performance in distinguishing between true positive and false negatives.
% This paper proposes an \gls{elc}  \gls{ll} algorithm with an evidential classifier . \gls{elc}  using evidential uncertainty, whilst . \gls{elc} 

This paper is organised as follows. Section II and Section III discuss related work and preliminary work, respectively. Section IV describes the methodology, including the \gls{elc} loss function and hyperparameters. Results are presented and discussed in Section V, and Section VI concludes the paper.

\section{Related Work}

\subsection{Lifelong Learning}\label{rw_ll}
Regularisation, constrained optimisation, replay, and architectural approaches have been used to tackle catastrophic forgetting. In the first 3 approaches, all tasks share the same \gls{nn} parameters and newer tasks are penalised in one form or another to prevent them from directly or indirectly increasing the loss on previous tasks. Regularisation approaches penalise changes in important weights for previous tasks (weight regularisation), or in the output of the model (function regularisation) \cite{cl_review2023}.

On the other hand, optimisation-based approaches define a constraint for the new task such that the previous loss is not increased. Replay-based approaches rely on using a representational sample of the old data while training for the new task: Experience replay relies on storing actual samples of the old data, whereas generative replay relies on an additional generative model to generate these samples \cite{cl_review2023}. 

Unlike previous approaches, architectural approaches isolate task-specific parameters, thereby eliminating inter-task interference. Architectural approaches may be applied to \gls{nn}s with fixed or expanding sizes. For a fixed-size \gls{nn}s, binary task-specific masks can be used to partition the \gls{nn} into task-specific parameters. For expanding \gls{nn}s, a combination of shared and task-specific layers (or subnetworks) can be used. Layers or subnetworks may be added to expanding \gls{nn}s when new tasks are needed \cite{cl_review2023}.

\subsection{Radar Waveform Classification}
Recent work done in radar signal classification includes modulation classification \cite{rc_wang2017, rc_jagannath2021, rc_Huang2023, rc_radyolo2024}, transfer learning \cite{jing2022radar_fewshot, xiao2020radar_fewshot, huang2025radar_fewshot}, and \gls{ll} (also known as incremental learning) \cite{xu2016_radar_il, han2023_radar_il}.

For modulation classification, \cite{rc_wang2017} convert \gls{iq} data into time-frequency images that are classified by a \gls{cnn}. \cite{rc_jagannath2021} proposes a framework that simultaneously classifies modulation and estimates signal characteristics. \cite{rc_Huang2023} converts \gls{iq} samples into time-frequency images to be classified by a \gls{cnn} and improve the approach in \cite{rc_jagannath2021} using an attention mechanism. These works show that \gls{cnn}s can achieve high classification accuracy. However, the problem of continually learning new waveforms is not addressed.

Other works focus on learning new waveforms using transfer learning, where knowledge of previously learned waveforms is reused to aid learning of new waveforms. \cite{xiao2020radar_fewshot} tackles the issue of over-fitting for small-scale radar datasets using transfer learning. Similarly, \cite{jing2022radar_fewshot} proposes an adaptive loss function that can transfer knowledge to learn difficult waveforms more efficiently. \cite{huang2025radar_fewshot} extends these ideas by proposing a self-supervised few-shot learning approach, capable of learning partially labelled new waveforms quickly. These works focus on efficiency of learning new waveforms by utilising prior knowledge. However, transfer learning on its own sacrifices prior knowledge and does not address catastrophic forgetting.

\cite{xu2016_radar_il} proposes an \gls{ll} algorithm for identifying radar emitters. This allows for new samples or features of old or new emitters to be incrementally learned. However, this approach relies on expert knowledge to extract features. Conversely, \cite{han2023_radar_il} proposes a \gls{nn} for incrementally learning new emitters. They propose a memory-based approach using point-exemplars (prototypes) of previously learnt emitters to mitigate (but not eliminate) catastrophic forgetting whilst learning new emitters.

The \gls{ll} approach used in this work is \gls{lps}. This approach combines incrementally learning new waveforms with transfer learning through adaptive masks. Unlike other approaches, mask-based \gls{ll} algorithms eliminate inter-task interference, thereby guaranteeing retention of prior knowledge. The details of \gls{lps} are expounded in Section \ref{subsection_ll}.

\subsection{Uncertainty Quantification and Selective Prediction}
\gls{uq} methods for \gls{ml} enable models to express uncertainty (or confidence) in predictions. Two types of uncertainties exist: aleatoric and epistemic. Aleatoric uncertainty is inherent in the data, caused by noise in the inputs or labels, and is irreducible. Epistemic uncertainty is due to the model's lack of knowledge (variation in its parameters) and can in principle be reduced by additional information \cite{al_epis_uncertainty}.

Selective Prediction in \gls{ml} allows models to abstain from making unreliable predictions based on a rejection cost. It allows for a trade-off between the proportion of accepted samples (coverage) and the recall over accepted samples (selective recall), based on a cost metric \cite{franc2023_sel_pred}. 
In this work, \gls{uq} is used as the cost metric for selective prediction. This enables the model to abstain from making predictions when its uncertainty is above a threshold. Section~\ref{sect:method_sel_pred} describes the algorithm used to find the optimal selection recall given a required coverage value, or vice-versa. 

\section{Preliminary Work}
\subsection{Lifelong Learning Algorithm}\label{subsection_ll}
A task is defined as a distinct learning episode that introduces new data which the model must learn sequentially without forgetting prior knowledge. Mask-based \gls{ll} algorithms partition a \gls{nn}'s weights for different tasks \cite{cl_review2023}. For each task $t$, the algorithm trains the \gls{nn} at full capacity, prunes the weights $\theta^t$ down to the pruning ratio $\alpha_t$, and then retrains the remaining weights for fine-tuning. Task-specific binary masks, $M^t$, for each layer $l$ and task $t$, $M^t = \{M_l^t\}_{l=1}^L \in \{0,1\}^m$, are generated from the remaining weights and are stored for future use. The remaining capacity of the \gls{nn} is $1- \bar{\alpha}_{t-1}$, where $\bar{\alpha}_{t-1}$ is the cumulative pruning ratio, $\sum_{i=0}^{t-1}\alpha_i$.

Before training subsequent tasks, gradients of parameters corresponding to non-zero elements in masks of previous tasks are set to zero, thereby excluding them from optimisation and preventing catastrophic forgetting. During inference task-specific masks are applied element-wise to the \gls{nn} parameters, nullifying parameters allocated to other tasks. This partitions the network such that parameters and masks associated with each task have disjoint supports from those of preceding tasks \eqref{eq_disjoint_mask}, where $\Theta^{t-1} = \sum_{i=1}^{t-1} \theta^i$ and $\operatorname{support}(f) = \{x \in X : f(x) \neq 0\}$.
\begin{subequations} \label{eq_disjoint_mask}
\begin{align}
\operatorname{support}(\theta^t) \;\;&\cap\; \operatorname{support}(\Theta^{t-1}) = \emptyset \label{eq_disjoint_weights} \\
\operatorname{support}(M^t) \;\;&\cap\;\; \operatorname{support}(\Theta^{t-1})  = \emptyset 
\end{align}
\end{subequations}

\gls{lps}\cite{lps} (the algorithm used in this work) takes the approach described in \eqref{eq_disjoint_mask} one step further, by allowing later tasks to learn `adaptive masks' during training. Unlike the original masks, adaptive masks are not required to uphold disjoint supports from masks of previous tasks, such that \eqref{eq_disjoint_mask} becomes $\operatorname{support}(M^t) \subseteq \operatorname{support}(\Theta^{t-1})$. That is to say: the parameters for previous tasks may be re-used, without modification, for the current task. The ratio of previous weights to be used, $\beta_i$, is a hyperparameter, and is utilised in the same way as the pruning ratio $\alpha_i$. 

The pruning method used for \gls{lps} is \gls{admm} \cite{admm}. Firstly, a desired sparse matrix, $Z$, which is initialised to $\theta^t$, is hard pruned according to $\alpha_t$. $\theta^t$ is then iteratively driven towards $Z$ according to the \gls{admm} loss, $\mathcal{L}_{\text{ADMM}}(\theta^t)$, in \eqref{eq_admm_weights}, where $U$ is the dual variable, $\rho$ is an \gls{admm} hyperparameter that penalises the difference between $\theta^t$ and $Z-U$. The same approach is used to learn the adaptive mask \eqref{eq_admm_mask}. The total \gls{admm} loss, $\mathcal{L}_{\text{ADMM}}(\theta^t, M^t)$, is given by \eqref{eq_admm}.
\begin{subequations} \label{eq_admm_loss}
\begin{align}
\mathcal{L}_{\text{ADMM}}(\theta^t) = \frac{\rho}{2} \left\| \theta^t - Z + U \right\|^2\label{eq_admm_weights} \\
\mathcal{L}_{\text{ADMM}}(M^t) = \frac{\tau}{2} \left\| M^t - Y + K \right\|^2\label{eq_admm_mask} \\
    \mathcal{L}_{\text{ADMM}}(\theta^t, M^t) = \mathcal{L}_{\text{ADMM}}(\theta^t) + \mathcal{L}_{\text{ADMM}}(M^t) \label{eq_admm} 
\end{align}
\end{subequations}

The total loss of the model, $\mathcal{L}_{\text{total}}$, is then the addition of $\mathcal{L}_{\text{ADMM}}$ to $\mathcal{L}_{\text{task}}$, which is the task-specific loss function used in the initial training phase (cross-entropy, in this case) \eqref{eq_lps_loss}. Thus, the loss function is optimised when the classifier is accurate ($\mathcal{L}_{\text{task}}\rightarrow0$) at the desired sparsity ($\mathcal{L}_{\text{ADMM}} \rightarrow0$). 
\begin{equation}
    \mathcal{L}_{\text{total}}(\theta^t, M^t) = \mathcal{L}_{\text{task}}(\theta^t) + \mathcal{L}_{\text{ADMM}}(\theta^t, M^t) \label{eq_lps_loss}
\end{equation}

\subsection{Evidential Classification} \label{subsect_evidential}
\subsubsection{Distance-based Belief Assignment}
Prototypes are point exemplars of each possible class. %Each prototype has a degree of membership to all classes, which determines its relative importance when predicting a class.
Classification of input data, $\textbf{x}$, occurs first by computing the distance-based support between $\textbf{x}$ and each prototype, defined as $s^i = \alpha^i~\phi(d^i)$, where $\alpha^i \in (0,1)$ is a scaling parameter and $\phi(d^i) \in (0,1)$ is a Gaussian Radial Basis Function. The mass associated with prototype $p^i$ in support of $\omega_q$ being the true class is computed by $m^i(\{\omega_q\}) = h_q^is^i$, where $h^i$ is the degree of membership of $p^i$ to class $\omega_q$ such that $\sum_{q=1}^Mh_q^i=1$. The mass functions are then aggregated using Dempster's Rule
% : $\textbf{m}=\bigoplus_{i=1}^M m^i$ 
\cite{ds_deno2000}.

\subsubsection{\gls{ds} Layer}
The evidential classifier proposed in \cite{ecnn-ds} takes a feature vector from a deep learning classifier and feeds it into a \gls{ds} layer, which converts it into a vector of mass functions, $\textbf{m}= \{m(\omega_1), \ldots, m(\omega_M), m(\Omega) \}$ . The mass functions within $\textbf{m}$, $m(\omega_i)$ for $i \in \{1,\ldots, M+1\}$, represent the mass of evidence strictly in support of $\omega_i$ being the true label, or lack thereof ($m(\Omega)$).

\subsubsection{Utility Layer} \label{sect_util}
The expected utility layer takes $\textbf{m}$ as an input and outputs the normalised expected utilities of acts. An act, $f_{\omega_i} \in \mathcal{F}$, denotes the assignment of a sample to class $\omega_i$, where $\mathcal{F} = \{f_{\omega_1}, \ldots, f_{\omega_M}\}$ is the set of all acts. 
% The desirability of the consequences of an act is referred to as its expected utility. 
The utility of an act's consequence reflects its desirability.
For a sample with a true label $\omega_j$, the utility of act $f_{\omega_i}$ is represented by $u_{ij}$. The expected utility, $\mathbb{E}_{m,\nu}$, of all acts \eqref{eq_ds_exputil} is computed from the expected minimum utility \eqref{eq_ds_minutil} and expected maximum utility \eqref{eq_ds_maxutil}, where $\Omega = \{\omega_1,\ldots, \omega_M\}$ is the set of all classes and $\nu \in[0,1]$ represents the pessimism of the classifier \cite{ecnn-ds}. For example, for $A = \{\omega_1, \omega_2\}$ and a true label $\omega_j = \omega_2$, $\nu$ controls how optimistic the classifier should be when guessing between $\omega_1$ and $\omega_2$, given it is confident that $\omega_j \subseteq A$. Thus, as $\nu \rightarrow1$, $\mathbb{E}_{m,\nu}(f_A)$ \eqref{eq_ds_exputil} tends towards the expected minimum utility, $\underline{\mathbb{E}}_m(f_A)$ \eqref{eq_ds_minutil}, and vice-versa.
\begin{subequations} \label{eq_ds_exputil_all}
\begin{align}
    \mathbb{E}_{m,\nu}(f_A) = \nu\; \underline{\mathbb{E}}_m(f_A) + (1 - \nu) \; \overline{\mathbb{E}}_m(f_A) \label{eq_ds_exputil} \\
    \underline{\mathbb{E}}_m(f_A) = \sum_{A \subseteq \Omega} m(A) \min_{\omega_j \in A} u_{ij} \label{eq_ds_minutil} \;\; \;\;\; \;\;\; \;\ \\
    \overline{\mathbb{E}}_m(f_A) = \sum_{A \subseteq \Omega} m(A) \max_{\omega_j \in A} u_{ij} \label{eq_ds_maxutil} \;\; \;\;\; \;\;\; \;\
\end{align}
\end{subequations}

The maximum of the expected utilities, $\mathbb{E}_{\nu}(f_{B})$, is given by \eqref{eq_maxutil}. It follows from here that epistemic uncertainty, $ \operatorname{u_{epistemic}}$, can be expressed as in \eqref{eq_ds_eu}, given normalised utilities \cite{ecnn-ds}.
% The act, $A^*$, that maximises $\mathbb{E}_{m,\nu}(f_{B})$ is given by \eqref{eq_A}, and the maximum of the expected utilities, $\mathbb{E}_{\nu}(f_{B})$, by \eqref{eq_maxutil}. It follows from here that epistemic uncertainty can be equated to $1-\mathbb{E}_\nu(f_A)$, given normalised utilities \cite{ecnn-ds}.
% \begin{equation}
% A^* = \arg\max_{\emptyset \neq A \subseteq \Omega} \mathbb{E}_{m,\nu}(f_A) \label{eq_A}
% \end{equation}
\begin{equation}
    \mathbb{E}_\nu(f_A) = 
    % \mathbb{E}_{m,\nu}(f_{A^*}) \label{eq_maxutil}
    \max_{\emptyset \neq A \subseteq \Omega} \mathbb{E}_{m,\nu}(f_A) \label{eq_maxutil}
\end{equation}
\begin{equation}
    \operatorname{unc_{epistemic}} =1-\mathbb{E}_\nu(f_A) \label{eq_ds_eu}
\end{equation}

\subsubsection{Evidential Loss Function}
The loss function \eqref{eq_l_ecnn} \cite{ecnn-ds} is similar to the standard $\text{CrossEntropy}$ loss, where $y_k \in \{0,1\} $ is $1$ when $\omega_k$ is the true label. As the epistemic uncertainty tends to $0$ (as $\mathbb{E}_\nu(f_{\omega_k}) \rightarrow 1$), the loss is minimised when the prediction is correct ($y_k =1$) and is maximised when the prediction is wrong ($y_k =0$); as the epistemic uncertainty tends to $1$ (as $\mathbb{E}_\nu(f_{\omega_k}) \rightarrow 0$), the loss is minimised when the prediction is wrong ($y_k =0$) and maximised when the prediction is correct ($y_k =1$). Thus, optimal loss occurs when the prediction is correct and the classifier is completely confident, or when the prediction is wrong and the classifier is completely uncertain.
\begin{equation}
    \mathcal{L_\text{DS}} = - \sum_{i=1}^{M} y_i \log \mathbb{E}_\nu(f_{\omega_i}) + (1 - y_i) \log \left(1 - \mathbb{E}_\nu(f_{\omega_i})\right) \label{eq_l_ecnn}
\end{equation}

\subsection{Bayesian Classification and Shannon Entropy} \label{subsect_bayes}
In Bayesian \glspl{nn}, model parameters are treated as random variables to capture uncertainty in predictions. A Bayesian linear layer replaces deterministic weights with probabilistic ones, such as \( w_i \sim \mathcal{N}(\mu_i, \sigma_i^2) \), enabling sampling-based inference. The predictive distribution is given by
\begin{align}
    p(y \mid x, \mathcal{D}) &= \int p(y \mid x, w)\, p(w \mid \mathcal{D})\, dw \nonumber \\
    &\approx \frac{1}{T}\sum_{t=1}^{T} p(y \mid x, w_t)
\end{align}
where \( w_t \) are Monte Carlo samples from the learned posterior $p(w \mid \mathcal{D})$, and $\mathcal{D}$ is the dataset. Variation across these samples reflects uncertainty in the model's parameters \cite{al_epis_uncertainty}.

Given this predictive distribution, total predictive uncertainty, $\operatorname{unc_{total}}$, is quantified using the Shannon entropy \cite{al_epis_uncertainty}: 
\begin{equation}
    H[y \mid x, \mathcal{D}] = -\sum_{c} p(y=c \mid x, \mathcal{D}) 
    \log p(y=c \mid x, \mathcal{D})
\end{equation}
Aleatoric uncertainty, $\operatorname{unc_{aleatoric}}$, is computed by taking the expectation of the weights over $H[y \mid x, \mathcal{D}]$:
\begin{equation}
    \operatorname{unc_{aleatoric}}=\mathbb{E}_{p(w \mid \mathcal{D})}[H[y \mid x, w]]
\end{equation}
This isolates the epistemic uncertainty by removing variation in the model parameters. Thus, epistemic uncertainty, $\operatorname{u_{epistemic}}$, is computed as:
\begin{equation}
    \operatorname{unc_{epistemic}}= \operatorname{unc_{total}} - \operatorname{unc_{aleatoric}}
\end{equation}
% \begin{align}
%     H[y \mid x, \mathcal{D}] &= 
%     }_{\text{Epistemic}} \nonumber \\
%     &\quad + \underbrace{}_{\text{Aleatoric}}
% \end{align}
% The epistemic term captures model uncertainty due to limited data, while the aleatoric term represents inherent data noise. Their sum constitutes the total predictive uncertainty .

\section{Methodology} \label{sect_methodology}

\subsection{Lifelong Learning and Uncertainty Quantification}
\gls{elc} and \gls{blc} integrate with \gls{lps} by replacing the final linear layer of the backbone ResNet with evidential or Bayesian layers, respectively. \gls{lps} applies \gls{ll} through mask-based partitioning of the ResNet. The task-specific loss $\mathcal{L}_{task}(\theta^t)$ \eqref{eq_lps_loss} is $\text{Cross-Entropy}$ except for \gls{elc}.

For \gls{elc}, a \gls{kl} divergence term, $D_{\text{KL}}$, is added to the evidential loss in \eqref{eq_l_ecnn}. 
This encourages \gls{elc} to be less confident when incorrect by penalising the \gls{kl} divergence between the utility vector ($\textbf{u}$) and a uniform distribution, encouraging a more even spread of utilities.
A soft-gate $w = u_{\text{max}}\cdot(1-u_{true\_class})$ is used to scale $D_\text{KL}$. $w$ is 0 for correct, confident predictions (when $u_{\text{max}}-u_{\text{true\_class}}=1$), and 1 for confident errors (when $u_{\text{max}}=1$, $u_{\text{true\_class}}=0$).
%(0 when $u_{\text{max}}=u_{\text{true\_class}}=1$; 1 when $u_{\text{max}}-u_{\text{true\_class}}=1$, for $u\in[0,1]$) is used to apply $D_{\text{KL}}$ based on confidence. 
\gls{elc}'s loss is thus given by \eqref{eqn:loss_elc}, where $\lambda_\text{KL}=10$ (annealed from 0). 
%A cosine scheduler is applied to $\lambda_\text{KL}$ ($0$-$5$) to avoid over-penalising early training. 
\begin{equation}
    \mathcal{L}_\text{ELC} = \mathcal{L}_\text{DS}+ w\cdot\lambda_{\text{KL}} \cdot  D_{\text{KL}} \label{eqn:loss_elc}
\end{equation}

% Line

% Line

% Line

% Line

% Line

% Line

% Line

% Line

% Line

% test

\subsection{Selective Prediction and Uncertainty Thresholding}\label{sect:method_sel_pred}

Selective prediction is employed to assess the model's ability to quantify and act upon predictive uncertainty. 
Each sample is assigned an uncertainty score $u(x)$, derived from its output distribution (i.e., Shannon entropy or evidential uncertainty). 
A confidence threshold $\tau$ is varied across the range of uncertainty values. 
Predictions with $u(x) \le \tau$ are accepted (trusted), whereas those with $u(x) > \tau$ are rejected. 
This simulates a decision system that issues a prediction only when confidence is sufficient, thus reducing the likelihood of unreliable outputs.  

For each threshold value, the selective recall (the recall computed over accepted samples) and the coverage (the proportion of accepted samples) are obtained using \eqref{eq:selacc}.
\begin{equation}
\text{Selective Recall}(\tau) =
\frac{\sum_i \ind\!\left\{\,u(x_i) \le \tau,\ \hat{y}_i = y_i\,\right\}}
     {\sum_i \ind\!\left\{\,u(x_i) \le \tau\,\right\}}
\label{eq:selacc}
\end{equation}
The complement of selective recall defines the selective risk \cite{franc2023_sel_pred}. 
Sweeping $\tau$ produces a risk--coverage curve that characterises the trade-off between prediction reliability and prediction coverage. This approach enables a principled evaluation of uncertainty calibration by emphasizing coverage-dependent reliability rather than overall recall alone.

\subsection{Datasets Description and Preprocessing}

\subsubsection{Data Preprocessing}\label{sect_preproc}
Where data are not in \gls{iq} format, it is first mixed down to baseband using Euler's identity, $y(t) = x(t)\exp(j2\pi f_ct)$, with $x(t)$ as the original signal and $f_c$ the carrier frequency. The resulting complex samples ($i_1 + jq_1, \ldots, i_n + jq_n$) are interleaved as $[i_1, q_1, \ldots, i_n, q_n]$ and low-pass filtered with a cut-off frequency $f_{\text{cutoff}} = f_{\text{BW}} / 2$, where $f_{\text{BW}}$ is the signal bandwidth. The signal is then downsampled by $M = f_s / (2 f_{\text{cutoff}})$. Processed samples are standardised and reshaped into 2D arrays of width 1024. Each row forms an input to the ResNet, labelled by the transmitter ID (\gls{rff}) or signal type (radar). For RadNIST, a 50\% overlapping sliding window is used to capture longer waveforms.

\subsubsection{Radar Datasets}
i) Synthetic Radar (RadNIST)~\cite{synth_radar} contains 
% in the 3.55–3.7~GHz band 
five types of pulses corresponding to radar working modes \cite{synth_radar}, sampled at 10~MS/s. It contains unmodulated (P0N) and frequency/angle modulated (Q3N) pulses, varying in phase-coding (Barker, Frank, P1-4/x, Zadoff-Chu) and bandwidth.
ii) Synthetic RadChar~\cite{rc_Huang2023} contains radar pulses sampled at 3.2~MS/s from five modulation schemes: coherent pulse train, barker code, polyphase barker code, frank code and linear frequency modulation). In both datasets, pulse parameters vary by pulse width, pulse repetition interval, and pulses per burst. 
%Both datasets include noise-only samples, which outnumber pulse-containing samples by $11\times$, and SNRs of $-20\operatorname{dB}$ to $18\operatorname{dB}$.
% In total, 50{,}000 pulses of 512~IQ samples are used.

\subsubsection{Communications Datasets}

i) Drone Remote Control (DRC)~\cite{dataset_drc} comprises signals from 17 drone controllers operating at 2.4~GHz with 10~MHz bandwidth, sampled at 20~GS/s and a \gls{snr} of 25~dB. %, yielding 20{,}000 samples per waveform and roughly 1000 waveforms per device. 
ii) LoRa~\cite{dataset_lora} contains 870~MHz radio frequency transmissions from 10 transmitters, with 125~kHz bandwidth, 50~dB \gls{snr}, and 1~MS/s sampling rate. %, yielding 800 signals of 2048~IQ samples per device after preprocessing.
% Data from 10 transmitters and a single receiver are used, yielding 800 signals of 2048~IQ samples per device after preprocessing.
iii) Identical USRP~\cite{ExposingRFF} includes transmissions from 20 identical \gls{usrp} devices sending identical WiFi packets, captured at 20~MS/s.%, yielding 13{,}000 waveforms of 512~IQ samples.

\subsection{Hyperparameters}
For the \gls{rff} datasets (DRC, LoRa, USRP), both homogeneous and heterogeneous task formations are used. In the homogeneous case, each task classifies a unique subset of devices from the same dataset (e.g., Task 1 and 2 classify devices 0–4 and 5–9, respectively). In the heterogeneous case, each task corresponds to a different dataset (e.g., Task 1: DRC, Task 2: LoRa). Radar datasets use only heterogeneous tasks due to the lack of a feasible homogeneous partition; these are prefixed with ``Mixed''.

Hyperparameters are set as follows: DRC, USRP, and Mixed RFF are split into 3 tasks; LoRa is split into 2 tasks; and Mixed Radar includes a task each for RadNIST and RadChar. Tasks of the same dataset contain roughly equal device counts. The adaptive mask ratio, $\beta$, is 0.9 for homogeneous tasks, and 0.2 and 0.1 for Mixed RFF and Mixed Radar, respectively. The pruning ratio, $\alpha$, equals the reciprocal of the number of tasks, except for Mixed RFF, where $\alpha = [0.15, 0.5, 0.35]$ for DRC, LoRa, and USRP tasks. Evidential hyperparameters are $\nu = 0.9$ and 20 prototypes per class. Training runs for a total of 100-150 epochs. The \gls{nn} input size is 1024.

\makeatletter
\setlength{\@fptop}{80pt} % default is 0pt; increase to push float down
\makeatother
\begin{table*}[ht]
    \centering
    \caption{Comparison of \gls{st}, Bayesian and \gls{elc} Recall ($\%$) on Tasks, Grouped by Dataset.}
    \label{tab:combined_results}
    \begin{tabular}{p{0.8cm} l ||
        >{\centering\arraybackslash}p{1.0cm}
        >{\centering\arraybackslash}p{1.2cm}
        >{\centering\arraybackslash}p{1.3cm} |
        >{\centering\arraybackslash}p{1.0cm} 
        >{\centering\arraybackslash}p{1.2cm} |
        >{\centering\arraybackslash}p{1cm} 
        >{\centering\arraybackslash}p{1.2cm} |
        >{\centering\arraybackslash}p{0.8cm} 
        >{\centering\arraybackslash}p{1.2cm} 
        }
        \toprule
        \textbf{Dataset} & \textbf{Task} 
        & \textbf{ST Linear} & \textbf{ST Bayesian}  & \textbf{ST Evidential} 
        & \textbf{LPS Task} & \textbf{LPS Task Avg.} 
        & \textbf{BLC Task} & \textbf{BLC Task Avg.}
        & \textbf{ELC Task} & \textbf{ELC Task Avg.} \\ 
        \hline \midrule 
        \multirow{2}{0.8cm}{\textbf{Mixed Radar}}
            & RadNIST & \multirow{2}{*}{87.0} & \multirow{2}{*}{88.3} & \multirow{2}{*}{85.1}
                      & 94.6 & \multirow{2}{*}{90.2} 
                      & 95.3 & \multirow{2}{*}{90.2} 
                      & 95.2 & \multirow{2}{*}{\textbf{90.8}} \\
            & RadChar & & & & 85.9 & & 85.0 & & 86.4 & \\
        \midrule
        \multirow{3}{*}{\textbf{DRC}} 
            & Devices 0-4 & \multirow{3}{*}{97.0} & \multirow{3}{*}{92.7} & \multirow{3}{*}{96.9}
                     & 98.3 & \multirow{3}{*}{97.9}
                     & 99.6 & \multirow{3}{*}{96.4}
                     & 99.9 & \multirow{3}{*}{\textbf{99.0}} \\
            & Devices 5-9 & & & & 99.8 & & 95.7 & & 97.5 & \\
            & Devices 10-14 & & & & 95.6 & & 93.8 & & 99.5 & \\
        \midrule
        \multirow{2}{*}{\textbf{LoRa}} 
            & Devices 0-4 & \multirow{2}{*}{81.1} & \multirow{2}{*}{83.9}  & \multirow{2}{*}{91.4}
                     & 90.0 & \multirow{2}{*}{90.6}  
                     & 90.9 & \multirow{2}{*}{91.2} 
                     & 95.7 & \multirow{2}{*}{\textbf{96.0}} \\
            & Devices 5-9 & & & & 91.1 & & 91.5 & & 96.2 & \\
        \midrule
        \multirow{3}{*}{\textbf{USRP}} 
            & Devices 0-4 & \multirow{3}{*}{74.2} & \multirow{3}{*}{75.5} & \multirow{3}{*}{71.5}
                     & 87.9 & \multirow{3}{*}{78.5} 
                     & 90.0 & \multirow{3}{*}{\textbf{83.0}} 
                     & 81.4 & \multirow{3}{*}{76.9} \\
            & Devices 5-9 & & & & 75.0 & & 82.3 & & 72.6 &  \\
            & Devices 10-14 & & & & 72.6 & & 76.8 & & 76.8 &  \\
        \midrule
        \multirow{3}{0.8cm}{\textbf{Mixed RFF}} 
            & DRC    & \multirow{3}{*}{87.5} & \multirow{3}{*}{62.4} & \multirow{3}{*}{67.4}
                     & 100.0 & \multirow{3}{*}{90.9} 
                     & 99.8 & \multirow{3}{*}{71.3} 
                     & 100.0 & \multirow{3}{*}{\textbf{92.0}} \\
            & LoRa   & & & & 92.2 & & 73.1 & & 90.0 & \\
            & USRP   & & & & 80.5 & & 41.1 & & 86.0 & \\
        \bottomrule
    \end{tabular}
\end{table*}

\section{Results}
\subsection{Overview}
The first set of results (Table~\ref{tab:combined_results}) compares the recall of 4 approaches. The first is \gls{st} which employs a single training phase (no \gls{ll}). \gls{st} models use the backbone ResNet with varied final layers: Linear, Bayesian and Evidential. \gls{st} approaches have the same architecture as their \gls{ll} equivalent approach (\gls{lps}, \gls{blc} and \gls{elc}), respectively.
The LL approaches incorporate the \gls{lps} \gls{ll} algorithm, with their differences being the final layers (see Section~\ref{sect_methodology}). The evaluation metric chosen is recall, defined as $\text{True Positives/(True Positives}+\text{False Negatives})$. The recall of each LL approach is assessed on individual tasks and then averaged for comparison against its \gls{st} equivalent. Selective prediction is evaluated on radar datasets using a global rejection threshold per dataset, and performance is then analysed by SNR. The resulting rejection ratio is therefore not constrained to be uniform across SNR levels.
% Selective prediction is evaluated on radar datasets as a function of \gls{snr}. Because rejection is driven by a global uncertainty threshold rather than SNR-stratified selection, the proportion of rejected samples differs across SNR levels.

% \gls{st} models are trained on the respective entire dataset at once. In contrast, the \gls{lps} approach partitions the datasets into multiple tasks, such that each task learns to classify a unique subset of devices. The average task recall is then compared to the \gls{st} recall to evaluate the performance of the \gls{lps} approach.

% the \gls{lps} approach is evaluated on each dataset and benchmarked against a \gls{st} baseline (\gls{lps} section of Table~\ref{tab:combined_results}). 
% Similarly, the \gls{elc} approach is evaluated against an \gls{st} \gls{ec} baseline (\gls{elc} section of Table~\ref{tab:combined_results}). 
\subsection{Discussion}
% \subsection{Comparison of Accuracy Across Datasets and Tasks}
Table~\ref{tab:combined_results} shows that the average task recall of the \gls{ll} models always outperform their \gls{st} equivalent; most significantly for \gls{blc} and \gls{elc} on Mixed \gls{rff} (+8.9\% and +24.6\%). \gls{elc} achieves higher average task recall on all datasets (+0.6\% to +20.7\%), except for on \gls{usrp} (-6.1\%). 

\begin{figure}[ht]
    \centering
    \begin{subfigure}[b]{0.45\textwidth}
        \centering
        \includegraphics[width=0.9\linewidth]{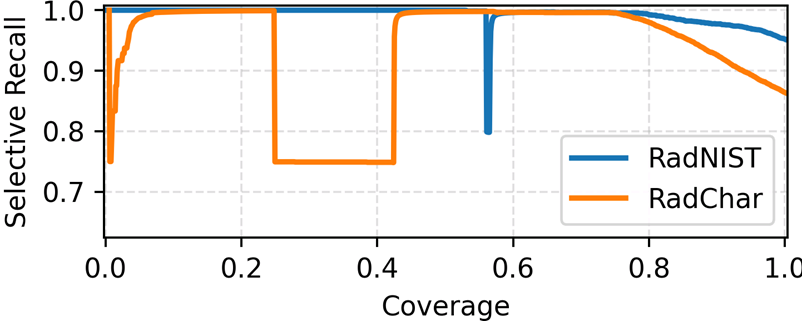}
        % \caption{\textbf{\gls{elc} on the Radar Datasets:} Labels 0-5 are
        % from the RadNIST dataset, and 6-11 are from RadChar.}\label{fig:ecnn-radars}
        \caption{Using evidential epistemic uncertainty (ELC).}\label{fig_cov_elc}
    \end{subfigure}
    \hfill
    \begin{subfigure}[b]{0.45\textwidth}
        \centering
        \includegraphics[width=0.9\linewidth]{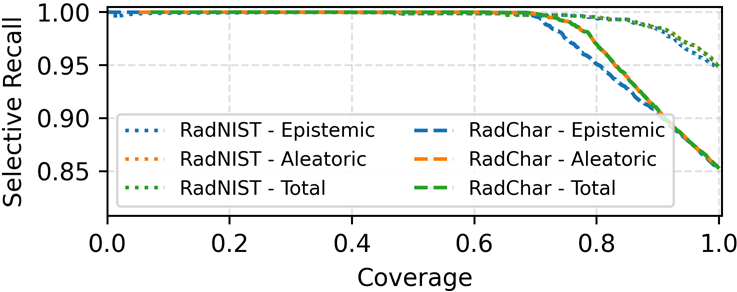}
        % \caption{Comparison of accuracy and uncertainty as SNR of testing data varies, for RadNIST and RadChar.}
        \caption{Using aleatoric, epistemic and total uncertainties (BLC).}
        \label{fig_cov_blc}
    \end{subfigure}
    \caption{Trade-off between selective recall and coverage on radar datasets of mixed SNR samples ($-20\operatorname{dB}$ to $18\operatorname{dB}$).}
    \label{fig:cov_acc}
\end{figure}

Note from Section~\ref{sect:method_sel_pred} that selective prediction offers a trade-off between selective recall and coverage, based on an uncertainty metric. Shannon entropy (for \gls{blc}) computes epistemic, aleatoric and total uncertainties, whereas \gls{elc} computes only epistemic uncertainty. Fig.~\ref{fig:cov_acc} shows that the selective recall--coverage trade-off varies significantly between datasets and models. Both plots show a general trend of increasing selective recall as coverage decreases. However, \gls{elc}'s trade-off (Fig.~\ref{fig_cov_elc}) is more fluctuant than \gls{blc}'s (Fig.~\ref{fig_cov_blc}). 
%Where the plots in Fig.~\ref{fig_cov_elc} stop before $0\%$ coverage indicates all remaining samples exhibit the same uncertainty level. 
Fig.~\ref{fig_cov_blc} shows that using aleatoric or total uncertainty has a marginal benefit over epistemic uncertainty for \gls{blc}.

% RadChar, sees a steep decrease in recall as the coverage increases beyond $90\%$. On the other hand, the RadNIST dataset saw almost no reduction in recall for up to $95\%$ coverage. Furthermore, Fig.~\ref{fig_cov_blc} shows a marginal benefit of using epistemic uncertainty over aleatoric or total uncertainty for \gls{blc}, which have overlapping graphs (orange and green). 

\begin{figure}[ht] 
    \centering
    \begin{subfigure}[b]{0.45\textwidth}
        \centering
        \includegraphics[width=1\linewidth]{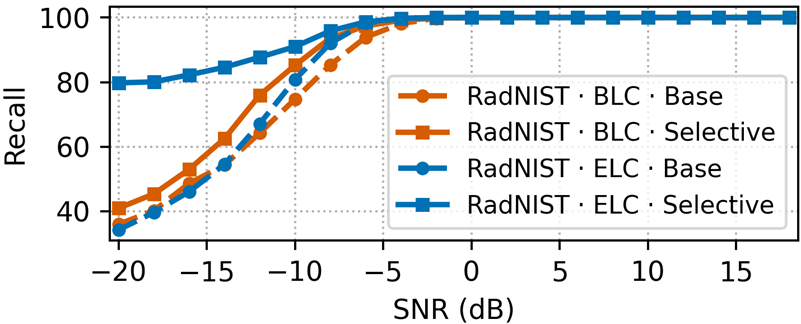}
        \caption{RadChar Dataset at $80\%$ Coverage.}
        \label{fig:snr_w/wo_sel_pred_radchar}
    \end{subfigure}
    \hfill
    \begin{subfigure}[b]{0.45\textwidth}
        \centering
        \includegraphics[width=1\linewidth]{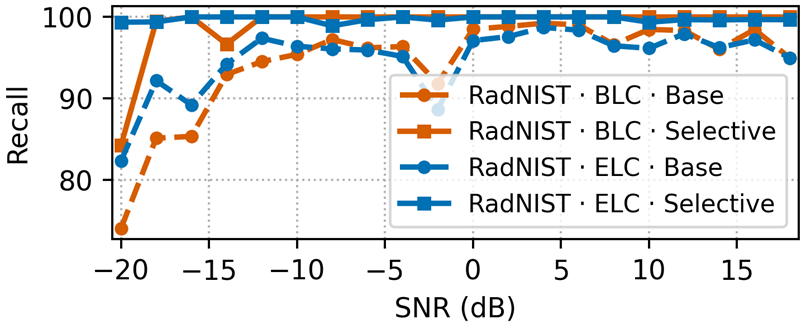}
        \caption{RadNIST Dataset at $80\%$ Coverage.}
        \label{fig:snr_w/wo_sel_pred_radnist}
    \end{subfigure}
    \caption{Recall with (selective) and without (base) selective prediction as a function of SNR, at $2\operatorname{dB}$ increments.}
    \label{fig:snr_w/wo_sel_pred}
\end{figure}

To quantify the benefit of using selective prediction, Fig.~\ref{fig:snr_w/wo_sel_pred} compares recall with and without selective prediction at an \gls{snr} range of $-20$-$18\operatorname{dB}$, for each of the radar datasets. 

Results vary significantly by dataset and model. On the RadChar dataset, Fig.~\ref{fig:snr_w/wo_sel_pred_radchar} shows that \gls{elc} benefits significantly, up to $+46\%$, at low \glspl{snr}; \gls{blc} only sees a $5$-$10\%$ improvement. Overall, \gls{elc}'s selective recall ($95.0\%$) outperforms \gls{blc}'s by $+7.3\%$.
On the RadNIST dataset, Fig.~\ref{fig:snr_w/wo_sel_pred_radnist} shows that \gls{elc} benefits up to $+17\%$ at low \glspl{snr}; \gls{blc} sees up to $15\%$ improvement. Overall, \gls{elc}'s selective recall ($99.7\%$) marginally outperforms \gls{blc}'s by $+0.7\%$.
On both datasets, \gls{elc} outperforms \gls{blc} at low \glspl{snr}.%, especially in selective recall.

% On the RadNIST dataset (Fig.~\ref{fig:snr_w/wo_sel_pred_radnist}), both models see improvement of up to $10$-$15\%$ and their selective recall approaches $100\%$.
%, with the improvement decreasing as selective recall approaches $100\%$. 
% Despite that, \gls{elc} outperforms \gls{blc} by $10$-$12\%$ at $-20~\text{dB}$ and has more consistent recall across \glspl{snr}.
%Interestingly, \gls{blc}'s selective recall rises steadily with \gls{snr} and marginally outperforms \gls{elc} at $-10$-$0\operatorname{dB}$.

% This demonstrates higher prediction reliability using selective prediction on both datasets.
% Both \gls{blc} and \gls{elc} see an improvement in average recall. However, \gls{elc} sees a significant an increase of up to $20\%$ and $50\%$ on the RadNIST (Fig.~\ref{fig:snr_w/wo_sel_pred_radnist}) and RadChar (Fig.~\ref{fig:snr_w/wo_sel_pred_radchar}) datasets respectively, which significantly outperform \gls{blc}.

\begin{figure}
    \centering
    \includegraphics[width=0.9\linewidth]{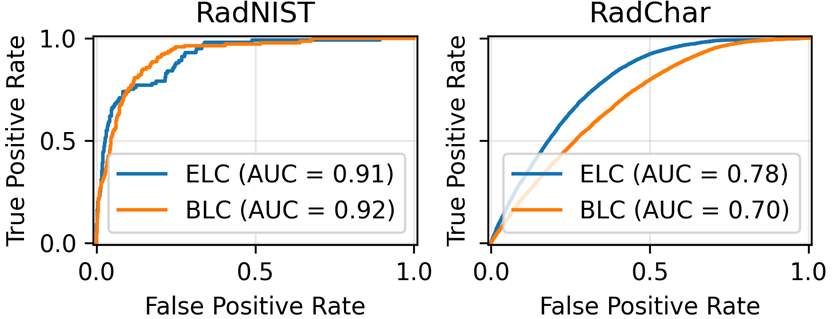}
    \caption{ROC plot for ELC and BLC at $\le-10\operatorname{dB}$ SNR.}
    \label{fig:unc_roc}
\end{figure}

To investigate the difference between \gls{elc} and \gls{blc} at $\le$$-10\operatorname{dB}$ \gls{snr}, uncertainty scores from both models are used to predict the correctness of their predictions. The ROC curves in Fig.~\ref{fig:unc_roc} show that both models are similar on the RadNIST dataset (within $1\%$). On RadChar, however, \gls{elc} is much more capable of distinguishing between true and false positives relative to \gls{blc} ($+12\%$), indicating a stronger correlation between uncertainty and correctness.


\section{Conclusion}
% In this work, we propose an evidential lifelong classifier by integrating Lifelong Learning with evidential classification. We demonstrate that the Lifelong Learning approach consistently outperforms the conventional Single-Task baseline. The evidential variant achieves slightly higher accuracy while also providing epistemic uncertainty. This uncertainty is valuable for i) balancing trade-offs between true and false positives, and ii) guiding targeted data collection or improvements in feature representation. Finally, we discuss the limitations of evidential classification, including increased training/inference time and larger model size.

\glsresetall % See this command
This work investigates the cross-section between \gls{ll} and decision making based on uncertainty quantification (selective prediction) for applications in the RF domain. This approach is evaluated on an \gls{elc}, which expresses uncertainty using evidence theory, and \gls{blc}, which expresses uncertainty using Shannon entropy. Results show that \gls{ll} variants consistently outperform Single Task training. For RF Fingerprinting and radar pulse classification \gls{elc} outperforms benchmarks by 0.2\%-20.7\%, except in one case (-6.1\%). 

This work also evaluates selective prediction on radar pulse datasets, to simulate uncertainty-based decision making. Results show that, at $80\%$ coverage, selective prediction improves confidence in \gls{elc}'s predictions at low SNRs, demonstrating effective rejection of unreliable predictions. \gls{elc} not only outperforms \gls{blc} in base recall, but also benefits more from selective prediction on average. This indicates that evidential uncertainty provides a stronger correlation between uncertainty and correctness of prediction. It must be noted that selective prediction does not improve the model; however, it improves trustworthiness of predictions by expressing ignorance.

\bibliography{references}

% \printglossaries

\end{document}

%% file: acronyms.tex
\newacronym{mfr}{MFR}{Multi-Functional Radar}
\newacronym{ai}{AI}{Artificial Intelligence}
\newacronym{sjr}{SJR}{Signal-to-Jamming Ratio}
\newacronym{rl}{RL}{Reinforcement Learning}
\newacronym{drl}{DRL}{Deep Reinforcement Learning}
\newacronym{eme}{EME}{Electromagnetic Environment}
\newacronym{ecm}{ECM}{Electromagentic Countermeasures}
\newacronym{rf}{RF}{Radio Frequency}
\newacronym{ew}{EW}{Electromagnetic Warfare}
\newacronym{cew}{CEW}{Cognitive Electromagnetic Warfare}
\newacronym{es}{ES}{Electromagnetic Surveillance}
\newacronym{mab}{MAB}{Multi-Armed Bandits}
\newacronym{sac}{SAC}{Soft-Actor Critic}
\newacronym{ddqn}{DDQN}{Double-Depth Q-Network}
\newacronym{dqn}{DQN}{Deep Q-Network}
\newacronym{ql}{QL}{Q-Learning}
\newacronym{a2c}{A2C}{Advantage-Actor Critic}
\newacronym{marl}{MARL}{Multi-Agent RL}
\newacronym{ppo}{PPO}{Proximal Policy Optimisation}
\newacronym{pri}{PRI}{Pulse Repetition Interval}
\newacronym{uav}{UAV}{Unmanned Aerial Vehicle}
\newacronym{per-ddqn}{PER-DDQN}{Prioritised Experience Replay DDQN}
\newacronym{dstl}{DSTL}{Defence Science and Technology Laboratory}
\newacronym{elint}{ELINT}{Electronic Intelligence}
\newacronym{tx-rx}{Tx-Rx}{transmitter-receiver}
\newacronym{cf}{CF}{Carrier Frequency}
\newacronym{bw}{BW}{Bandwidth}
\newacronym{pw}{PW}{Pulse Width}
\newacronym{lps}{LPS}{Learn-Prune-Share}
\newacronym{admm}{ADMM}{Alternating Direction Method of Multipliers}
\newacronym{rn}{ResNet}{Residual Network}
\newacronym{relu}{ReLu}{Rectified Linear Unit}
\newacronym{rc}{RC}{Remote Control}
\newacronym{usrp}{USRPs}{Universal Software Radio Peripherals}
\newacronym{nist}{NIST}{National Institute of Standards and Technology}
\newacronym{awgn}{AWGN}{Additive White Gaussian Noise}
\newacronym{snr}{SNR}{Signal-to-Noise Ratio}
\newacronym{ds}{DS}{Dempster-Shafer}
\newacronym{lora}{LoRa}{Long Range Radio}
\newacronym{ll}{LL}{Lifelong Learning}
\newacronym{cnn}{CNN}{Convolutional Neural Networks}
\newacronym{drc}{DRC}{Drone Remote Controllers}
\newacronym{rx}{Rx}{receiver}
\newacronym{rff}{RFF}{RF Fingerprinting}
\newacronym{iot}{IoT}{Internet of Things}
\newacronym{ml}{ML}{Machine Learning}
\newacronym{iq}{IQ}{In-phase and Quadrature}
\newacronym{mtl}{MTL}{Multi-Task Learning}
\newacronym{cl}{CL}{Continual Learning}
\newacronym{lpf}{LPF}{Low-Pass Filter}
\newacronym{comms}{Comms.}{communications}
\newacronym{st}{ST}{Single Task}
\newacronym{elc}{ELC}{Evidential Lifelong Classifier}
\newacronym{nn}{NN}{Neural Network}
\newacronym{ec}{EC}{Evidential Classifier}
\newacronym{blc}{BLC}{Bayesian Lifelong Classifier}
\newacronym{uq}{UQ}{Uncertainty Quantification}
\newacronym{kl}{KL}{Kullback-Leibler}